\documentclass[aps,prb,superscriptaddress,reprint,amssymb,nobibnotes,longbibliography]{revtex4-2}
\usepackage[english]{babel}
\usepackage{graphicx}
\usepackage{subfigure}
\usepackage{verbatim}
\usepackage{dsfont}
\usepackage[version=3]{mhchem}
\usepackage{chemfig}
\usepackage{cancel}
\usepackage{amsmath}
\usepackage{xcolor}
\usepackage{upgreek}
\usepackage{bm}
\usepackage{amssymb}
\usepackage{braket}
\usepackage{microtype}
\usepackage{textcomp}
\usepackage{hyperref}

\makeatletter
\newsavebox{\@brx}
\newcommand{\llangle}[1][]{\savebox{\@brx}{\(\m@th{#1\langle}\)}%
\mathopen{\copy\@brx\kern-0.5\wd\@brx\usebox{\@brx}}}
\newcommand{\rrangle}[1][]{\savebox{\@brx}{\(\m@th{#1\rangle}\)}%
\mathclose{\copy\@brx\kern-0.5\wd\@brx\usebox{\@brx}}}
\makeatother

\begin{document}

\title{Boltzmann-constrained extraction of spin splitting and momentum relaxation in $d$-wave altermagnets}

\author{Y. X. Gao}
\affiliation{College of Physics, Nanjing University of Aeronautics and Astronautics, Nanjing 211106, China}
\affiliation{Key Laboratory of Aerospace Information Materials and Physics (NUAA), MIIT, Nanjing 211106, China}

\author{Z. W. Fan}
\affiliation{School of Engineering, Newcastle University, Newcastle upon Tyne, NE1 7RU, UK}

\author{Q. S. Yao}
\email{yaoqiushi@nuaa.edu.cn}
\affiliation{College of Physics, Nanjing University of Aeronautics and Astronautics, Nanjing 211106, China}
\affiliation{Key Laboratory of Aerospace Information Materials and Physics (NUAA), MIIT, Nanjing 211106, China}

\author{Y. D. Ji}
\email{jiyanda@nuaa.edu.cn}
\affiliation{College of Physics, Nanjing University of Aeronautics and Astronautics, Nanjing 211106, China}
\affiliation{Key Laboratory of Aerospace Information Materials and Physics (NUAA), MIIT, Nanjing 211106, China}

\author{H. Geng}
\email{genghao@nuaa.edu.cn}
\affiliation{College of Physics, Nanjing University of Aeronautics and Astronautics, Nanjing 211106, China}
\affiliation{Key Laboratory of Aerospace Information Materials and Physics (NUAA), MIIT, Nanjing 211106, China}

\date{\today}

\begin{abstract}
Altermagnets exhibit spin-split electronic structure without requiring spin-orbit coupling, but transport measurements generally mix intrinsic spin splitting with extrinsic scattering. We examine this identifiability problem for a two-dimensional $d$-wave altermagnet within a unified semiclassical framework spanning ballistic to diffusive transport. The spin-dependent Fermi-surface anisotropy produces a pronounced size effect, where vastly different longitudinal velocities cause the two spin channels to exhibit markedly different effective relaxation lengths within the same device geometry. However, the altermagnetic coupling $\alpha$ and the momentum relaxation time $\tau_0$ strongly compensate each other in longitudinal conductance, creating a severe parameter degeneracy. To lift this degeneracy, we formulate a physics-informed neural network (PINN) to act as a differentiable Boltzmann solver that strictly enforces contact injection, local particle conservation, and global current continuity. Driven by sparse conductance spectra, this neural solver leverages the Fermi-level dependence of transport to unlock the coupled parameters simultaneously, achieving sub-percent accuracy even under moderate measurement noise. These results show that combining the Fermi-level dependence of transport with strict physical constraints provides a robust route to separating spin splitting from scattering in altermagnetic conductors.
\end{abstract}

\maketitle

\section{Introduction}

Altermagnetism has emerged as a distinct form of collinear magnetic order in which crystal symmetries generate momentum-dependent spin splitting even in the absence of net magnetization and relativistic spin-orbit coupling \cite{RN144,RN143,RN137,RN157}.  This combination of antiferromagnetic compensation and spin-polarized band structure makes altermagnets attractive for spin transport, because their spin texture can affect charge motion without the stray fields characteristic of ferromagnets.  A central question for transport studies is therefore not only whether altermagnetic spin splitting is present, but how its microscopic strength can be separated from ordinary scattering processes in a device measurement.

This separation is difficult because the most accessible observables are macroscopic averages over a spin- and angle-dependent Fermi surface.  Spectroscopic probes such as angle-resolved photoemission can be complicated by multidomain averaging in candidate materials \cite{RN20}, while transport signals are also affected by contact geometry, impurity scattering, and competing spin-charge conversion mechanisms \cite{RN21}.  In a Boltzmann description, the altermagnetic coupling changes the velocity distribution on the Fermi surface, whereas the relaxation time controls how rapidly injected carriers lose memory of the contacts.  These two effects can compensate in the longitudinal conductance: increasing the spin splitting may enhance the contribution of fast carriers, while reducing the relaxation time suppresses their propagation.  A reliable extraction of microscopic parameters must therefore resolve a genuine kinetic inverse problem rather than fit a single conductance curve.

The semiclassical Boltzmann equation provides the natural bridge between microscopic band dynamics and mesoscopic transport \cite{RN146,RN147,RN11,RN7,RN158}.  In particular, a unified semiclassical formulation that interpolates continuously from the diffusive to the ballistic regime has been established for charge and anomalous transport~\cite{RN11,RN7,RN158}, in which the nonequilibrium distribution is determined self-consistently by a local equilibrium enforcing particle conservation together with directional carrier injection at the contacts.  We adopt and extend this framework to the spin-split Fermi surface of a $d$-wave altermagnet.  For a strongly anisotropic Fermi surface, however, direct numerical solution requires angular and spatial discretization of a nonlocal self-consistent integral equation.  This is manageable for a forward calculation but becomes cumbersome when repeated many times in a two-parameter inverse search.  Conversely, purely data-driven regression can reproduce measured conductances without enforcing particle conservation or current continuity, and therefore may identify parameter sets that do not correspond to a physically admissible steady state \cite{RN23,RN133,RN135,RN28,RN24}.  The issue is thus physical as much as numerical: the inverse problem should be regularized by the same kinetic equations that define the transport observable.

Here we reformulate the problem as a Boltzmann-constrained extraction of the altermagnetic coupling $\alpha$ and relaxation time $\tau_0$ from sparse Fermi-level-dependent conductance data.  We use a minimal $d$-wave altermagnetic Hamiltonian to derive the spin-resolved Boltzmann transport equation, the local particle-conservation condition, and the conductance functional.  A physics-informed neural network (PINN) provides a differentiable representation of the distribution function, which is then optimized under the Boltzmann residual, contact boundary conditions, and current conservation; in the inverse mode, $\alpha$ and $\tau_0$ are optimized as physical parameters rather than treated as labels.  This framework is used to address three questions.  First, how accurately can the steady-state electrochemical-potential profile and conductance be reconstructed across diffusive, crossover, and ballistic regimes?  Second, what spin-resolved kinetic texture is produced by the $d$-wave altermagnetic Fermi surface?  Third, under what conditions can the conductance compensation between $\alpha$ and $\tau_0$ be lifted?

The main physical finding is that the altermagnetic Fermi-surface anisotropy, when coupled with the finite device size, produces a pronounced spin-dependent kinetic contrast that is visible in the reconstructed nonequilibrium distribution but largely compressed in the scalar conductance.  Mapping the conductance error in the $(\alpha,\tau_0)$ plane exposes the corresponding compensation valley.  Adding the local Boltzmann equation and global continuity constraint converts this valley into a well-conditioned inverse calculation within the parameter range studied, enabling simultaneous extraction of spin splitting and scattering from sparse conductance spectra.

The remainder of this paper is organized as follows.  Section~\ref{sec:formulation} introduces the minimal $d$-wave altermagnetic model and derives the spin-resolved Boltzmann equation, the local particle-conservation condition, and the conductance functional.  Section~\ref{sec:inversion} formulates the differentiable constrained solver together with its forward and inverse losses.  Section~\ref{sec:results} presents the results: a forward benchmark across the diffusive, crossover, and ballistic regimes; the spin-resolved kinetic texture of the altermagnetic Fermi surface; the single-parameter extraction of $\alpha$; and the joint inversion of $(\alpha,\tau_0)$ that lifts the conductance compensation.  Section~\ref{sec:conclusion} concludes.

\section{Boltzmann formulation for a two-dimensional $d$-wave altermagnet}\label{sec:formulation}

We consider a two-dimensional mesoscopic channel of length $L_x$ connected to ideal source and drain reservoirs.  The low-energy electronic structure is described by the minimal single-particle Hamiltonian \cite{RN137,RN144,RN157}
\begin{equation}
\hat{H}(\mathbf{k}) = \frac{\hbar^2 k^2}{2m}\hat{I} + \alpha(k_x^2 - k_y^2)\hat{\sigma}_z,
\end{equation}
where $m$ is the effective mass, $\alpha$ is the altermagnetic coupling, and $\hat{\sigma}_z$ is defined with respect to the N\'eel-vector axis.

The symmetry content of this model can be made precise in the language of spin space groups (SSG)~\cite{RN160,RN161}, with the generators identified via the Qsymm algorithm~\cite{RN159}. Since $[\hat{H}(\mathbf{k}),\hat{\sigma}_z]=0$, the spin along the N\'eel axis is conserved [SSG element $[E\,\|\,C_{\infty z}^S]$], and the isotropic kinetic term $\frac{\hbar^2 k^2}{2m}\hat{I}$ is trivial in spin space. The altermagnetic term $\alpha(k_x^2 - k_y^2)\hat{\sigma}_z$ is protected by the defining altermagnetic generator $g=[C_{4z}^R\,\|\,C_{2x}^S]$, which acts on the Hamiltonian as $U\hat{H}(\mathbf{k})U^{-1}=\hat{H}(R\mathbf{k})$ with the spin operation $U=\hat{\sigma}_x$ and the $\pi/2$ momentum rotation
\begin{equation}
R=R(\pi/2)=\begin{pmatrix} 0 & -1 \\ 1 & 0 \end{pmatrix},\qquad (k_x,k_y)\rightarrow(-k_y,k_x).
\end{equation}
The rotation $R$ sends $(k_x^2 - k_y^2)\rightarrow -(k_x^2 - k_y^2)$, while $U=\hat{\sigma}_x$ maps $\hat{\sigma}_z\rightarrow -\hat{\sigma}_z$ (since $\hat{\sigma}_x\hat{\sigma}_z\hat{\sigma}_x^{-1}=-\hat{\sigma}_z$); the two sign changes cancel and the term is invariant. The same decoupling of spatial and spin operations forbids a conventional ferromagnetic coupling $(k_x^2 + k_y^2)\hat{\sigma}_z$, whose spatial part carries no compensating sign change, enforcing the unconventional $d_{x^2-y^2}$-wave momentum dependence of the spin splitting.

The two spin branches have dispersions
\begin{equation}
\varepsilon_{\uparrow\downarrow}(\mathbf{k}) = \frac{\hbar^2}{2m}(k_x^2+k_y^2) \pm \alpha(k_x^2-k_y^2).
\end{equation}
For $|\alpha|<\hbar^2/(2m)$ the Fermi contours are closed ellipses with orthogonal principal axes for the two spin branches.  Introducing $k_x=k\cos\theta$ and $k_y=k\sin\theta$, we write
\begin{equation}
\varepsilon_\sigma(k,\theta)=A_\sigma(\theta)k^2,\qquad
A_\sigma(\theta)=\frac{\hbar^2}{2m}+s_\sigma\alpha\cos 2\theta,
\end{equation}
where $s_\uparrow=+1$ and $s_\downarrow=-1$.  The Fermi wave vector is $k_{F\sigma}(\theta)=\sqrt{E_F/A_\sigma(\theta)}$, and the Fermi-surface area element reduces to
\begin{equation}
 dk_xdk_y=\frac{1}{2A_\sigma(\theta)}d\varepsilon d\theta.
\end{equation}
This angular representation is the form used below for both the kinetic equation and conductance integral.

In a weak longitudinal bias, the steady-state Boltzmann equation for spin branch $\sigma$ in the relaxation-time approximation is
\begin{equation}\label{eq.bte_steady_reframed}
 v_{\sigma x}\frac{\partial f_\sigma}{\partial x}
 - eE_xv_{\sigma x}\frac{\partial f_0}{\partial\varepsilon_{\mathbf{k}}}
 =-\frac{f_\sigma-\bar f}{\tau_0},
\end{equation}
where $v_{\sigma x}$ is the longitudinal group velocity, $f_0$ is the global equilibrium Fermi-Dirac distribution, $\bar f$ is the local equilibrium distribution, and $\tau_0$ is the momentum relaxation time.  We assume a semiclassical regime with well-defined quasiparticles and slowly varying fields \cite{RN158}.  The local equilibrium distribution is
\begin{equation}
\bar f(\mathbf{k},x)=\left[\exp\left(\frac{\varepsilon_\mathbf{k}-eV(x)-\mu(x)}{k_BT}\right)+1\right]^{-1},
\end{equation}
where $V(x)$ and $\mu(x)$ denote the electrostatic potential and chemical potential.  In linear response we expand
\begin{equation}
 f_\sigma(\mathbf{k},x)=f_0(\varepsilon_\mathbf{k})+
 \left(-\frac{\partial f_0}{\partial\varepsilon_\mathbf{k}}\right)
 [eV(x)+g_\sigma(\mathbf{k},x)],
\end{equation}
\begin{equation}
 \bar f(\mathbf{k},x)=f_0(\varepsilon_\mathbf{k})+
 \left(-\frac{\partial f_0}{\partial\varepsilon_\mathbf{k}}\right)
 [eV(x)+\bar g(x)],
\end{equation}
with $\bar g(x)=\mu(x)-E_F$.  Substitution into Eq.~(\ref{eq.bte_steady_reframed}) cancels the explicit electric-field term and gives the transport equation for the electrochemical-potential perturbation,
\begin{equation}\label{eq.bte_simplified_reframed}
 v_{\sigma x}\frac{\partial g_\sigma}{\partial x}
 =-\frac{g_\sigma-\bar g}{\tau_0}.
\end{equation}

The collision term conserves local particle number, so $\bar g(x)$ is fixed by
\begin{equation}
\sum_\sigma\int f_\sigma(\mathbf{k},x)d\mathbf{k}
=\sum_\sigma\int \bar f(\mathbf{k},x)d\mathbf{k}.
\end{equation}
At zero temperature, $-\partial f_0/\partial\varepsilon_\mathbf{k}=\delta(\varepsilon_\mathbf{k}-E_F)$, and the conservation condition becomes
\begin{equation}\label{eq.gbar_reframed}
\bar g(x)=
\frac{\sum_\sigma\int_0^{2\pi} \frac{g_\sigma(x,\theta)}{2A_\sigma(\theta)}d\theta}
{\sum_\sigma\int_0^{2\pi}\frac{1}{2A_\sigma(\theta)}d\theta}.
\end{equation}

Here a single local electrochemical potential $\bar g(x)$ is shared by both spin channels, rather than a spin-resolved $\bar g_\sigma(x)$. Although $[\hat{H}(\mathbf{k}),\hat{\sigma}_z]=0$ makes the two spin branches dynamically independent, the collision term drives the distribution toward a \emph{common} local equilibrium fixed by conservation of the \emph{total} carrier density [Eq.~(\ref{eq.gbar_reframed})]. This number-conserving local-equilibrium construction, together with the directional contact injection introduced below, is the central element of the unified semiclassical theory that bridges the diffusive and ballistic regimes~\cite{RN11,RN158}; the present formulation generalizes it to the two spin-split branches of the altermagnet. Such a unified description is appropriate for the two-terminal charge measurement considered here, in which non-magnetic contacts inject both spin species at the same electrochemical potential and no net spin accumulation is generated. A fully spin-resolved treatment with independent $\bar g_\sigma(x)$ would instead be required for magnetic contacts or spin-accumulation phenomena, which lie outside the present charge-transport analysis.

\begin{figure*}[htbp]
    \centering
    \includegraphics[width=0.9\linewidth]{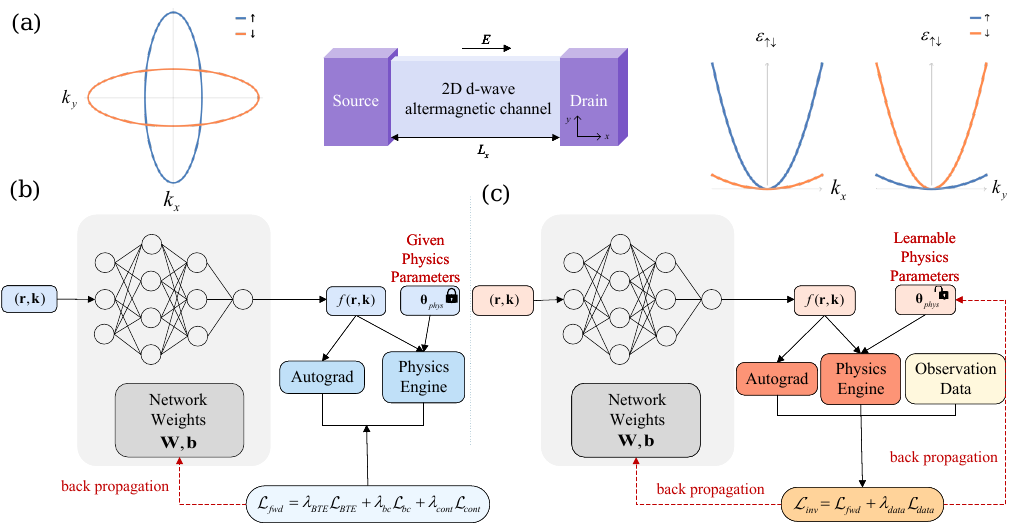}
    \caption{Boltzmann-constrained route from microscopic parameters to transport observables.  (a) A two-dimensional altermagnetic channel is driven by a small source-drain bias. The spin-resolved Fermi contours and schematic spin-split band dispersions along $k_x$ and $k_y$ illustrate the minimal dispersion $\varepsilon_{\uparrow\downarrow}(\mathbf{k})=\hbar^{2}(k_x^{2}+k_y^{2})/(2m)\pm\alpha(k_x^{2}-k_y^{2})$, with the Fermi contours shown for $\alpha=0.4$. The measured longitudinal conductance depends on both the altermagnetic coupling ($\alpha$) and momentum relaxation time ($\tau_0$).  (b) Forward mode: for fixed microscopic parameters, a differentiable representation of $g_\sigma(x,\theta)$ is optimized subject to the Boltzmann equation, directional contact injection, and current continuity.  (c) Inverse mode: the same constraints are retained, but $\alpha$ and/or $\tau_0$ are treated as unknown physical parameters and optimized using sparse conductance data $G(E_F)$.}
    \label{Fig1}
\end{figure*}

The ideal-contact boundary conditions are directional: right-moving carriers injected at $x=0$ satisfy $g_\sigma(0,\theta)=g_{\sigma L}=eV_L$ for $v_{\sigma x}>0$, while left-moving carriers injected at $x=L_x$ satisfy $g_\sigma(L_x,\theta)=g_{\sigma R}=eV_R$ for $v_{\sigma x}<0$ \cite{RN12}.  The corresponding formal solution of Eq.~(\ref{eq.bte_simplified_reframed}) is \cite{RN11,RN158}
\begin{widetext}
\begin{equation}
\begin{split}
 g_\sigma(\mathbf{k},x)=&\,\Theta(v_{\sigma x})\Bigg[g_{\sigma L}e^{-x/(v_{\sigma x}\tau_0)}
 +\int_0^x e^{-(x-\xi)/(v_{\sigma x}\tau_0)}
 \bar g(\xi)\frac{d\xi}{v_{\sigma x}\tau_0}\Bigg] \\
&+\Theta(-v_{\sigma x})\Bigg[g_{\sigma R}e^{-(L_x-x)/(|v_{\sigma x}|\tau_0)}
 +\int_x^{L_x} e^{-(\xi-x)/(|v_{\sigma x}|\tau_0)}
 \bar g(\xi)\frac{d\xi}{|v_{\sigma x}|\tau_0}\Bigg].
\end{split}
\end{equation}
\end{widetext}
Substituting this solution into Eq.~(\ref{eq.gbar_reframed}) yields a closed self-consistent integral equation,
\begin{equation}
 \bar g(x)=h(x)+\int_0^{L_x}K(x,\xi)\bar g(\xi)d\xi,
\end{equation}
with
\begin{widetext}
\begin{equation*}
\begin{split}
 &h(x)=\frac{1}{\sum_\sigma\int_0^{2\pi}\frac{1}{2A_\sigma(\theta)}d\theta}
\Bigg[\sum_\sigma\int_{v_{\sigma x}>0}g_{\sigma L}e^{-x/(v_{\sigma x}\tau_0)}\frac{d\theta}{2A_\sigma(\theta)}
+\sum_\sigma\int_{v_{\sigma x}<0}g_{\sigma R}e^{-(L_x-x)/(|v_{\sigma x}|\tau_0)}\frac{d\theta}{2A_\sigma(\theta)}\Bigg],\\
 &K(x,\xi)=\frac{1}{\sum_\sigma\int_0^{2\pi}\frac{1}{2A_\sigma(\theta)}d\theta}
\sum_\sigma\int_0^{2\pi}\frac{e^{-|x-\xi|/(|v_{\sigma x}|\tau_0)}}{|v_{\sigma x}|\tau_0}
\frac{d\theta}{2A_\sigma(\theta)}.
\end{split}
\end{equation*}
\end{widetext}
This equation makes the parameter-correlation problem explicit: both $\alpha$ and $\tau_0$ enter the decay kernel through the spin-dependent velocities and relaxation length.

The longitudinal current density is obtained by averaging the nonequilibrium perturbation over the Fermi surface,
\begin{equation}\label{eq.current_reframed}
\begin{split}
 j_x&=\frac{e}{(2\pi)^2}\sum_\sigma\int v_{\sigma x}
 \left(-\frac{\partial f_0}{\partial\varepsilon_\mathbf{k}}\right)
 g_\sigma(\mathbf{k},x)d^2\mathbf{k}\\
&=\frac{e}{h^2}\sum_\sigma\int_0^{2\pi}v_{\sigma x}
 \frac{g_\sigma(x,\theta)}{2A_\sigma(\theta)}d\theta.
\end{split}
\end{equation}
Equations~(\ref{eq.gbar_reframed}) and (\ref{eq.current_reframed}) imply $\partial j_x/\partial x=0$ in a source-free steady state.  The conductance is then
\begin{equation}
 G=\frac{j_xL_y}{V_L-V_R}.
\end{equation}

\section{Differentiable constrained inversion}\label{sec:inversion}

Figure~\ref{Fig1} summarizes the route from the minimal $d$-wave altermagnetic channel to the differentiable forward and inverse transport problems: panel (a) connects the spin-resolved Fermi contours and spin-split band dispersions to the source--drain geometry, while panels (b) and (c) distinguish the fixed-parameter forward mode from the parameter-learning inverse mode.

The integral formulation above can be solved by spatial and angular discretization, but the inverse problem requires the forward map $\{\alpha,\tau_0\}\mapsto G(E_F)$ to be evaluated and differentiated repeatedly.  We therefore employ a PINN to provide a differentiable representation of the spin-resolved distribution,
\begin{equation}
 \mathcal{N}_{\mathbf{W},\mathbf{b}}:(x,\theta,E_F,\sigma)\mapsto g_\sigma(x,\theta;E_F),
\end{equation}
where $\mathbf{W}$ and $\mathbf{b}$ are network weights and biases.  Automatic differentiation provides $\partial g_\sigma/\partial x$, while angular integrals are estimated from sampled collocation points.  The use of a neural representation is only a numerical device; the admissible solution space is defined by the Boltzmann residual, boundary injection, particle conservation, and current continuity.

For internal collocation points $\{x_i,\theta_j\}$, the transport residual is
\begin{equation}
\begin{split}
\mathcal{L}_{\mathrm{BTE}}=&\frac{1}{N_{\mathrm{col}}}\sum_{i,j}\sum_\sigma
\bigg|v_{\sigma x,j}\frac{\partial g_\sigma(x_i,\theta_j)}{\partial x}
+\frac{g_\sigma(x_i,\theta_j)-\bar g(x_i)}{\tau_0}\bigg|^2 .
\end{split}
\end{equation}
The directional contact loss is
\begin{equation}
\begin{split}
\mathcal{L}_{\mathrm{bc}}=&\frac{1}{N_{\mathrm{bc}}}\sum_j\sum_\sigma
\Big[\Theta(v_{\sigma x,j})|g_\sigma(0,\theta_j)-g_{\sigma L}|^2\\
&+\Theta(-v_{\sigma x,j})|g_\sigma(L_x,\theta_j)-g_{\sigma R}|^2\Big].
\end{split}
\end{equation}
To enforce the global source-free steady state we also penalize deviations from current continuity,
\begin{equation}
 \mathcal{L}_{\mathrm{cont}}=\frac{1}{N_x}\sum_{i=1}^{N_x}\left|\frac{\partial j_x(x_i)}{\partial x}\right|^2.
\end{equation}
The forward loss is
\begin{equation}
 \mathcal{L}_{\mathrm{fwd}}=\lambda_{\mathrm{BTE}}\mathcal{L}_{\mathrm{BTE}}
 +\lambda_{\mathrm{bc}}\mathcal{L}_{\mathrm{bc}}
 +\lambda_{\mathrm{cont}}\mathcal{L}_{\mathrm{cont}}.
\end{equation}

In the inverse calculation, the conductance data $G_{\mathrm{obs}}(E_{F,i})$ enter through
\begin{equation}
 \mathcal{L}_{\mathrm{data}}=\frac{1}{N_{\mathrm{data}}}\sum_{i=1}^{N_{\mathrm{data}}}
 |G_{\mathrm{pred}}(E_{F,i})-G_{\mathrm{obs}}(E_{F,i})|^2,
\end{equation}
so that
\begin{equation}
 \mathcal{L}_{\mathrm{inv}}=\mathcal{L}_{\mathrm{fwd}}+
 \lambda_{\mathrm{data}}\mathcal{L}_{\mathrm{data}}.
\end{equation}
The unknown physical parameters are updated together with the distribution representation, but they are constrained at every step by the same kinetic equations used in the forward problem.

\begin{figure*}[htbp]
    \centering
    \includegraphics[width=0.9\textwidth]{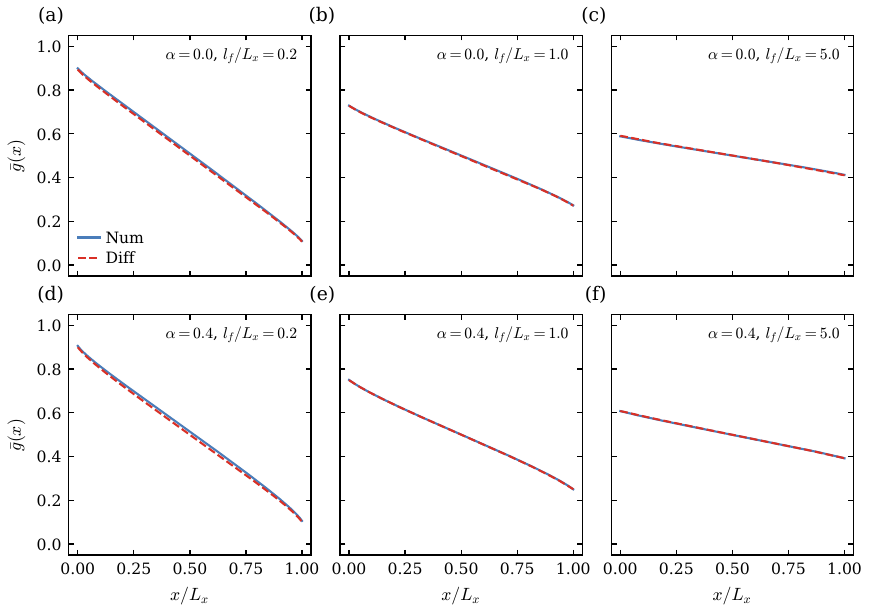}
    \caption{Local electrochemical-potential profile $\bar g(x)$ across diffusive, crossover, and ballistic regimes.  Blue solid curves are dense-discretization Boltzmann solutions and red dashed curves are differentiable-solver results.  (a)--(c) Normal metal, $\alpha=0$.  (d)--(f) $d$-wave altermagnet, $\alpha=0.4$.  From left to right, $l_f/L_x=0.2$, $1.0$, and $5.0$, with $E_F=1.0$.}
    \label{Fig2}
\end{figure*}

\begin{figure*}[htbp]
    \centering
    \includegraphics[width=0.95\linewidth]{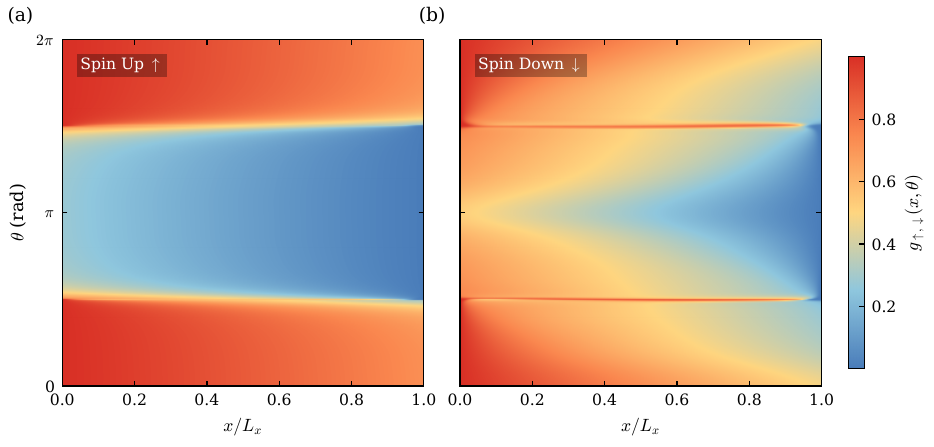}
    \caption{Spin-resolved nonequilibrium distributions in a $d$-wave altermagnetic channel.  The heat maps show (a) $g_\uparrow(x,\theta)$ and (b) $g_\downarrow(x,\theta)$ for $\alpha=0.4$, $E_F=1.0$, and $l_f/L_x=1.0$.  The contrasting textures reflect the spin-dependent longitudinal velocities produced by the altermagnetic Fermi-surface anisotropy.}
    \label{Fig3}
\end{figure*}

\begin{figure*}[htbp]
    \centering
    \includegraphics[width=\textwidth]{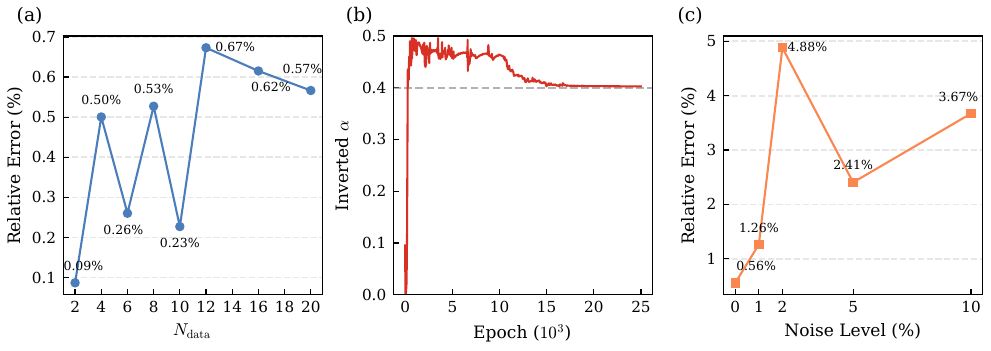}
    \caption{Single-parameter inversion of the altermagnetic coupling $\alpha$ with fixed $\tau_0$.  (a) Relative error versus the number of conductance data points $N_{\mathrm{data}}$.  (b) Representative optimization trajectory for $N_{\mathrm{data}}=8$ and target $\alpha=0.4$.  (c) Relative error under Gaussian noise in the conductance data; points are averaged over three independent runs.}
    \label{Fig4}
\end{figure*}

Unless otherwise noted, the network contains five hidden layers with 128 neurons per layer and hyperbolic tangent activations.  A bounded activation is used at the output to keep the perturbation within the numerical range set by the bias window.  The Adam optimizer is used for both network parameters and physical parameters.  For the two-parameter inversion shown below, the learning rates are $10^{-3}$ for $(\mathbf{W},\mathbf{b})$ and $5\times10^{-3}$ for $(\alpha,\tau_0)$, with cosine annealing and gradual activation of the loss weights.  The calculations were performed on a single NVIDIA GeForce RTX 4090 GPU.  These implementation details are not part of the physical model; they specify how the constrained optimization was carried out.

\section{Results and Discussion}\label{sec:results}

\subsection{Forward benchmark across transport regimes}

We first verify that the differentiable solver reproduces the Boltzmann solution before using it for parameter extraction.  The benchmark is a dense spatial-angular discretization of the self-consistent integral equation for $\bar g(x)$.  Because very small $|v_{\sigma x}|$ produces long collision kernels for carriers moving nearly transverse to the channel, the reference calculation uses a cutoff $|v_{\sigma x}|>0.02$ in the collision integral.  This benchmark provides the local electrochemical-potential profile and the conductance for normal ($\alpha=0$) and altermagnetic ($\alpha=0.4$) channels.

Figure~\ref{Fig2} compares the calculated $\bar g(x)$ for $l_f/L_x=0.2$, $1.0$, and $5.0$.  The three regimes interpolate from an almost linear diffusive profile to a weakly varying ballistic profile.  The agreement shows that the constrained representation captures both the local relaxation to $\bar g(x)$ and the boundary-injected nonequilibrium component.  Table~\ref{Tab1} gives the corresponding conductance comparison.  The relative conductance error remains below $0.14\%$ for all six cases.  The timing comparison is included only as a practical reference for the repeated inverse calculations; the essential point is that the forward map is differentiable with respect to $\alpha$ and $\tau_0$.

\begin{table}[htbp]
\caption{Forward benchmark against a dense-discretization Boltzmann solver.}
\label{Tab1}
\begin{ruledtabular}
\begin{tabular}{ccccccc}
$\alpha$ & $l_f/L_x$ & $G_{\mathrm{Num}}$ & $G_{\mathrm{Diff}}$ & Error (\%) & $T_{\mathrm{Num}}$ (s) & $T_{\mathrm{Diff}}$ (s) \\
\colrule
0.0 & 0.2 & 1.3367 & 1.3384 & 0.13 & 686.0129 & 510.9195 \\
0.0 & 1.0 & 3.3808 & 3.3761 & 0.14 & 685.6756 & 511.5833 \\
0.0 & 5.0 & 4.9423 & 4.9409 & 0.03 & 685.4178 & 511.4807 \\
0.4 & 0.2 & 2.1266 & 2.1265 & 0.00 & 686.8765 & 513.6723 \\
0.4 & 1.0 & 5.0716 & 5.0646 & 0.14 & 681.1062 & 511.9175 \\
0.4 & 5.0 & 7.3120 & 7.3048 & 0.10 & 685.7128 & 510.2357 \\
\end{tabular}
\end{ruledtabular}
\end{table}

\subsection{Spin-resolved kinetic texture of altermagnetic transport}

\begin{figure*}[htbp]
    \centering
    \includegraphics[width=\textwidth]{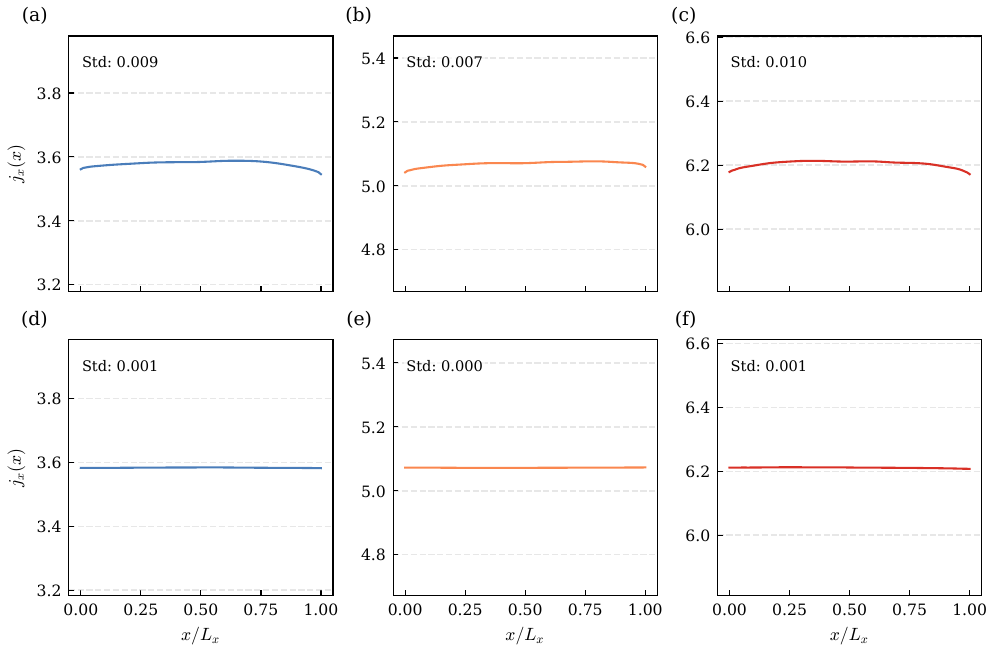}
    \caption{Role of current conservation in the inverse calculation.  (a)--(c) Reconstructed current density $j_x(x)$ without the current-continuity penalty ($\lambda_{\mathrm{cont}}=0$) at $E_F=0.5$, $1.0$, and $1.5$.  (d)--(f) Corresponding results with current continuity included ($\lambda_{\mathrm{cont}}=0.5$).  The suppressed spatial standard deviation demonstrates that global conservation removes nonphysical steady states allowed by local residual fitting alone.}
    \label{Fig5}
\end{figure*}

The scalar conductance compresses the microscopic kinetic structure.  The reconstructed phase-space distributions in Fig.~\ref{Fig3} show what is lost in that compression.  For $\alpha=0.4$, $E_F=1.0$, and $l_f/L_x=1.0$, the two spin branches have orthogonal Fermi-surface distortions and therefore different longitudinal velocity distributions.  In the spin-up channel, the larger longitudinal velocity leads to a long effective mean free path comparable to the device size, allowing the injected distribution to retain a sharp contact memory over a substantial fraction of the device.  In the spin-down channel, the smaller longitudinal velocity shortens the effective relaxation length and rapidly smooths the injected distribution toward local equilibrium.

This strong disparity in effective relaxation lengths---a manifestation of finite size effects in the presence of strong velocity anisotropy---is the microscopic origin of the parameter-extraction problem.  The altermagnetic coupling changes not only the magnitude of the conductance but also the distribution of relaxation lengths over the Fermi surface.  A transport inversion should therefore infer $\alpha$ from the kinetic evolution of $g_\sigma(x,\theta)$, not only from a scalar change in $G$.

\subsection{Single-parameter extraction from sparse conductance spectra}

We next test the simpler inverse problem in which $\tau_0$ is fixed and the altermagnetic coupling $\alpha$ is extracted from conductance values at selected Fermi levels.  The target is $\alpha_{\mathrm{true}}=0.4$.  Figure~\ref{Fig4}(a) shows that the relative error remains below $0.67\%$ when the number of conductance points is varied from 2 to 20.  The weak dependence on data volume indicates that the conductance points are not being used as independent labels for a black-box fit; they select a solution within a space already restricted by the Boltzmann equation and conservation laws.  Figure~\ref{Fig4}(b) shows a representative convergence trajectory for $N_{\mathrm{data}}=8$.  The parameter moves rapidly toward the target and then relaxes slowly as the kinetic residual and data loss are balanced.

Noise robustness is shown in Fig.~\ref{Fig4}(c).  With $N_{\mathrm{data}}=8$, Gaussian noise levels from $1\%$ to $10\%$ were added to the conductance data.  Even at $10\%$ noise, the mean relative error in $\alpha$ remains $3.67\%$.  This robustness should be interpreted as a consequence of kinetic regularization: noisy conductance values must still be compatible with a distribution satisfying the Boltzmann equation, contact injection, and continuity.

The current-continuity constraint is essential for this behavior.  Figure~\ref{Fig5} compares the reconstructed current density with and without the penalty on $\partial j_x/\partial x$.  Without the continuity constraint, boundary fitting can produce spatially varying currents even when the local residual is small.  Including the constraint suppresses these unphysical variations and restores the source-free steady state over the full device.  For inverse extraction, this prevents the optimizer from using a nonconserved current profile to compensate for an incorrect microscopic parameter.

We also tested different target values of $\alpha$, including weak-coupling cases close to the isotropic-metal limit.  For $\alpha=0.05$ the absolute error is $|\Delta\alpha|=0.0067$, showing that the approach remains sensitive to small Fermi-surface distortions in the present minimal model.

\subsection{Resolving the compensation between spin splitting and scattering}

\begin{figure}[htbp]
    \centering
    \includegraphics[width=\linewidth]{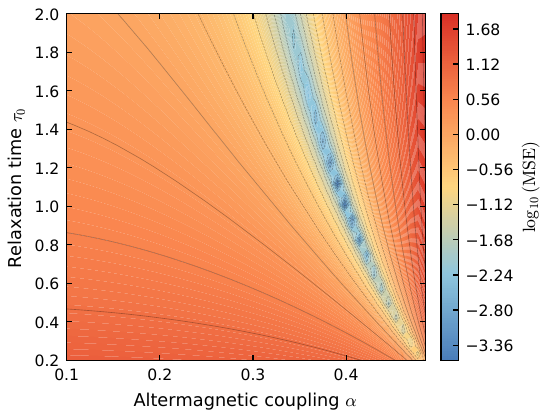}
    \caption{Conductance-error landscape in the two-parameter space.  The color scale shows $\log_{10}$ of the mean-squared error between trial conductance spectra and the target spectrum.  The narrow curved valley demonstrates compensation between the altermagnetic coupling $\alpha$ and relaxation time $\tau_0$.}
    \label{Fig6}
\end{figure}

In realistic transport characterization, neither $\alpha$ nor $\tau_0$ is known a priori.  Figure~\ref{Fig6} shows why the joint inversion is substantially harder than the single-parameter case.  The plotted quantity is the mean-squared difference between conductance spectra generated at a trial pair $(\alpha,\tau_0)$ and those at the target pair.  The low-error region forms a narrow curved valley rather than an isolated circular basin.  Along this valley, a larger altermagnetic spin splitting can be partly offset by shorter momentum relaxation, leaving a similar longitudinal conductance spectrum.  This is the transport compensation that must be resolved.

\begin{figure}[htbp]
    \centering
    \includegraphics[width=\linewidth]{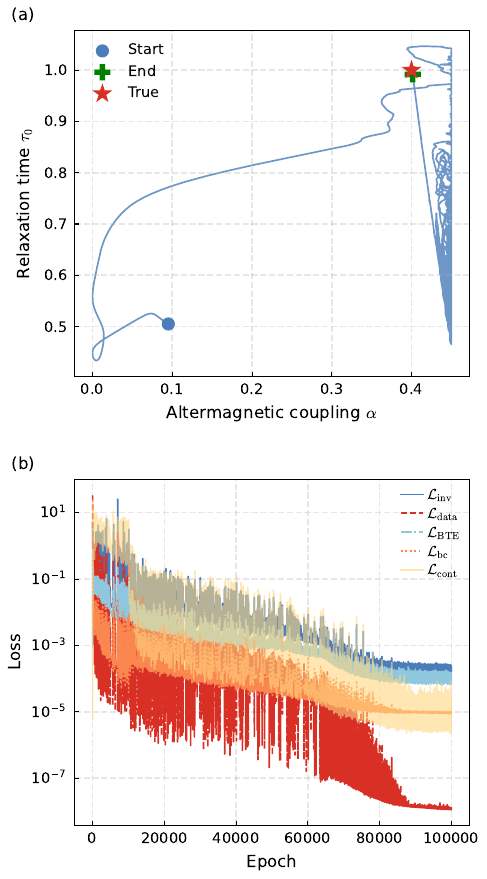}
    \caption{Joint inversion of $\alpha$ and $\tau_0$.  (a) Optimization trajectory in parameter space.  The path reflects the anisotropic error landscape: rapid approach in the spin-splitting direction is followed by slower relaxation along the scattering direction.  (b) Convergence of the inverse, data, Boltzmann, boundary, and continuity losses.}
    \label{Fig7}
\end{figure}

The Boltzmann-constrained inverse calculation uses $N_{\mathrm{data}}=12$ conductance points while retaining the kinetic and continuity losses.  Figure~\ref{Fig7}(a) shows a representative optimization path in the $(\alpha,\tau_0)$ plane.  The trajectory initially follows the shallow direction of the compensation valley, then turns toward the target as the distribution function becomes consistent with both the conductance data and the kinetic constraints.  The two parameters relax on different scales: $\alpha$, which controls the angular velocity structure, is pinned relatively early, whereas $\tau_0$, which controls the overall relaxation length, evolves more slowly along the valley.  Figure~\ref{Fig7}(b) shows the accompanying decrease of the loss components.

For the target values $\alpha=0.4$ and $\tau_0=1.0$, the extracted values in the noise-free case are $\alpha=0.4014$ and $\tau_0=0.9917$, corresponding to relative errors of $0.36\%$ and $0.83\%$.  With $1\%$ Gaussian noise added to the conductance data, the extracted values are $\alpha=0.4017$ and $\tau_0=0.9990$, with relative errors of $0.43\%$ and $0.10\%$.  The improvement in $\tau_0$ for this noisy realization should not be viewed as a general benefit of noise; rather, it indicates that moderate noise did not destabilize the constrained solution in the tested parameter range.  The relevant conclusion is that the joint extraction remains stable when the measured scalar conductance is perturbed.

\section{Conclusion}\label{sec:conclusion}

We have reformulated the characterization of a $d$-wave altermagnet as a Boltzmann-constrained inverse transport problem.  In the minimal model studied here, the altermagnetic coupling $\alpha$ controls the angular spin splitting and produces spin-dependent ballistic--diffusive propagation, while the relaxation time $\tau_0$ controls the decay of injected nonequilibrium carriers.  Because these effects compensate in longitudinal conductance, direct scalar fitting produces a narrow low-error valley in the $(\alpha,\tau_0)$ plane.

The differentiable Boltzmann solver used here provides two functions.  In forward mode it reproduces benchmark electrochemical-potential profiles and conductances over diffusive-to-ballistic regimes.  In inverse mode it allows $\alpha$ and $\tau_0$ to be optimized while enforcing the local kinetic equation, directional contact injection, local particle conservation, and global current continuity.  This converts sparse Fermi-level-dependent conductance data into a constrained kinetic reconstruction and enables the simultaneous extraction of spin splitting and scattering within the tested parameter range.

The present calculation is intentionally minimal.  It neglects spin-orbit coupling, multiband crossings, magnetic domains, coherent interference, and contact disorder, and the inverse tests use synthetic conductance data generated from the same semiclassical model.  These restrictions make the identifiability mechanism transparent, but they also define the next steps toward experimental use.  Extending the approach to material-specific band structures, finite-temperature broadening, multidomain averaging, and additional transport channels will determine how robustly altermagnetic spin splitting can be extracted from real device measurements.  More broadly, the physically constrained neural representation demonstrated here offers a rigorous paradigm for utilizing machine learning to resolve complex, ill-posed inverse problems across condensed matter physics.

\section*{Data Availability}
The data that support the findings of this study are available from the corresponding authors upon reasonable request. The source code is publicly available at GitHub: \url{https://github.com/GaoYX88/Boltzmann-Constrained-Altermagnet-Transport}.

\begin{acknowledgments}
This work was supported by the National Natural Science Foundation of China under Grant No. 12304068 (H.G.), the startup Fund of Nanjing University of Aeronautics and Astronautics Grant No. YAH24076 (H.G.). The computations are partially supported by the High Performance Computing Platform of Nanjing University of Aeronautics and Astronautics.
\end{acknowledgments}

\bibliography{ref}

\end{document}